\documentclass[aps,prb,preprint,groupedaddress,citeautoscript,nopacs]{revtex4}

\usepackage{graphicx}
\usepackage{color}
\usepackage{amsfonts}
\setcitestyle{super}

\begin{document}

\title{Quantum Criticality of Topological Phase Transitions in 3D Interacting Electronic Systems}

\author{Bohm-Jung Yang$^{1}$, Eun-Gook Moon$^{2}$, Hiroki Isobe$^{3}$ and Naoto Nagaosa$^{1,3}$}

\affiliation{$^1$ RIKEN Center for Emergent Matter Science (CEMS), Wako, Saitama 351-0198, Japan}

\affiliation{$^2$ Department of Physics, University of California, Santa Barbara, CA 93106, USA}

\affiliation{$^3$ Department of Applied Physics, University of Tokyo, Tokyo 113-8656, Japan}

\date{\today}

\begin{abstract}
{\bf
Topological phase transitions in condensed matters accompany
emerging singularities of the electronic wave function,
often manifested by gap-closing points in the momentum space.
In conventional topological insulators in three dimensions (3D),
the low energy theory near the gap-closing point
can be described by relativistic Dirac fermions coupled to the long range Coulomb interaction,
hence the quantum critical point of topological phase transitions
provides a promising platform to test the novel predictions
of quantum electrodynamics.
Here we show that a new class of quantum critical phenomena emanates
in topological materials breaking either the inversion symmetry or
the time-reversal symmetry.
At the quantum critical point, the
theory is described by the emerging low energy fermions, dubbed the anisotropic Weyl fermions,
which show both the relativistic and Newtonian dynamics simultaneously.
The interplay between the anisotropic dispersion and the Coulomb interaction
brings about a new screening phenomena distinct from
the conventional Thomas-Fermi screening in metals and logarithmic screening in Dirac fermions.
}
\end{abstract}

\maketitle
Fathoming quantum phases and the associated phase transitions is
one of the most significant problems in modern condensed matter physics.
Through the natural extension of classical order parameters,
the conventional symmetry breaking paradigm has been successfully
applied to quantum systems and
has deepened our understanding, especially of magnetism and superconductivity.\cite{subir,leon,cenke}
On the other hand, the recent theoretical advances in the study of
topological orders have provided solid theoretical grounds to predict exotic topological phases.
Due to the lack of the classical counterparts, the description of
these exotic topological phases and the phase transitions between them
is far beyond the realm of the conventional paradigm.\cite{xiaogang}
For instance, through the recent active studies of the systems
with strong spin-orbit coupling, a new topological semi-metal (SM),
dubbed the Weyl SM, is proposed in systems
such as A$_2$Ir$_2$O$_7$~\cite{Weyl_iridate, tokura, weyl} or topological insulator multilayers~\cite{Burkov}.
In Weyl SM, composed of several gapless points (Weyl points) on the Fermi energy $E_{F}$,
the bulk energy spectrum shows the 3D linear dispersion relation
around each Weyl point (WP).
Contrary to the conventional phases distinguished by
order parameters, the Weyl SM is characterized by the topological number, so-called the chirality
of WP whose nontrivial topological nature gives rise
to the ``Fermi arc'' surface states.
Since the chirality of each WP guarantees its stability,
the topological phase transition from the Weyl SM must accompany the pair-annihilation
of WPs.~\cite{weyl}
In this way, the nature of the Weyl SM and its topological phase transition
can be clearly described as long as the interaction between electrons is neglected.

\begin{figure*}[t]
\centering
\includegraphics[width=16 cm]{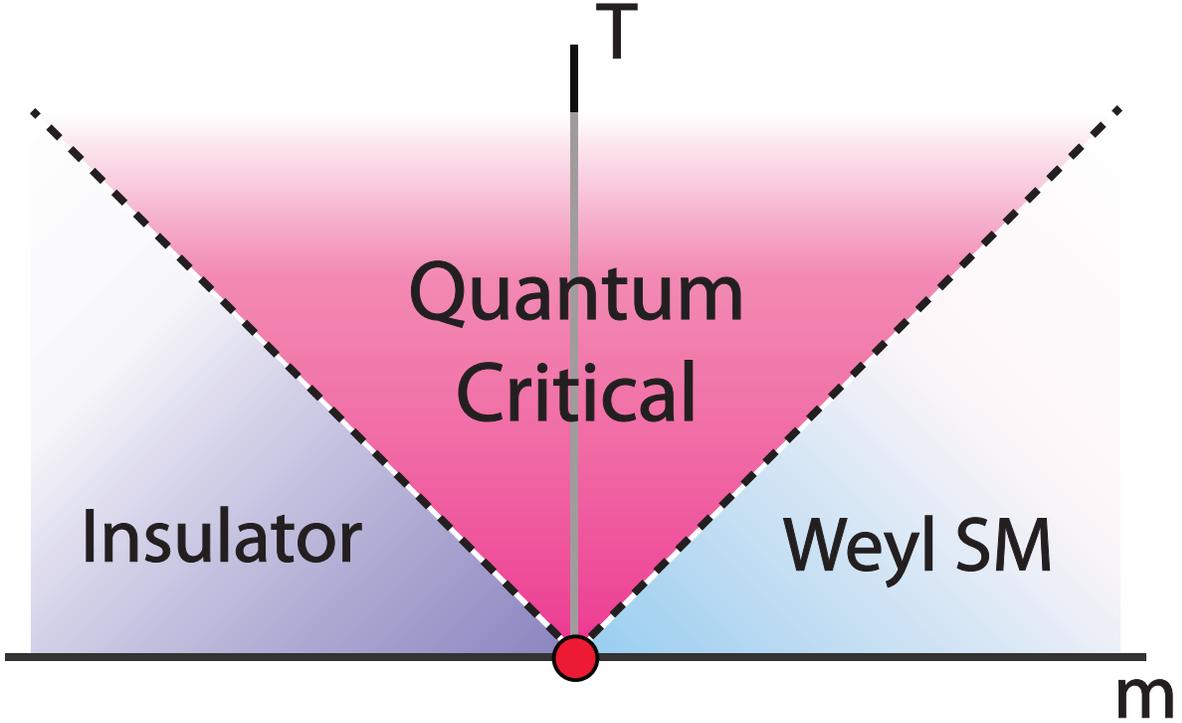}
\caption{
{\bf Generic phase diagram for topological phase transitions in 3D systems lacking
either the time-reversal symmetry or the inversion symmetry.}
Finite temperature phase diagram of the 3D topological
phase transition between an insulator and Weyl SM with
the tuning parameter $m$.
The AWF emerges at the quantum critical point with $m=0$.
By replacing the Weyl SM with a topological insulator,
the same phase diagram can describe the direct topological phase transition
between two insulators.
The finite temperature crossover is
described by the dotted lines which represent
energy gap in the case of the insulator while
they indicate the energy scale relevant to the linear dispersion
in Weyl SM. The three regimes show
characteristic behaviors in physical quantities. For example, the specific heat
shows $C_{\text{I}} \sim e^{-|m|/T}$ in the insulator, $C_{\text{SM}} \sim T^3$ in the Weyl SM,
and $C_{\text{QC}} \sim T^{5/2}$ in the quantum critical regime.
It is shown that the long range
Coulomb interaction is non-trivial in the quantum critical regime in sharp
contrast to the case of the Weyl SM where the interaction causes only
logarithmic corrections.}
\label{fig1}
\end{figure*}

However, the complete understanding of topological phases and related phase transitions
in interacting Weyl SM requires the careful consideration of electron correlation effects.
For instance, because of the relativistic nature of the electronic dispersion in Weyl SM,
the fermionic excitations coupled to the long range Coulomb interaction
induce logarithmic corrections to all physical quantities.~\cite{Hosur,Isobe1,Isobe2,Chakravarty}
Moreover, in the presence of the cubic and time reversal symmetries,
the interplay between the fermions and
the long range Coulomb interaction induces the non-Fermi liquid phase after the pair-annihilation
of WPs.~\cite{Luttinger1,Luttinger2,Moon}
In fact, in generic systems without high symmetries,
the pair-annihilation of WPs generates insulating phases
as shown in Fig.~\ref{fig1}.
Since the critical point is connected to the insulating phase,
the effective theory of the quantum phase transition should respect the zero-chirality condition.
Thus, the minimal model for the phase transition is the 3D anisotropic Weyl fermion (AWF)
whose dispersion is quadratic along one direction but linear along the other
two orthogonal directions.
It is worth to note that AWF can emerge ubiquitously in materials
breaking either the time-reversal symmetry or the
inversion symmetry, and mediate various topological phase transitions.~\cite{Yang,Murakami1,Murakami2}
For instance, the recent
theoretical~\cite{Yang,BiTeI1} and experimental~\cite{BiTeI2} studies on BiTeI
have shown that the same critical theory describes the direct topological phase
transition between the trivial and topological insulators.
Therefore for the universal description of topological quantum phase transitions
in generic topological materials, it is crucial to unveil the intriguing quantum criticality
of the 3D AWF coupled to the long range Coulomb interaction,
which is the main subject in this study.

The main finding of this work is as follows.
Most importantly,
we find that the anisotropic dispersion of AWF induces exotic screening effects.
Namely, significant screening appears along the direction with linear dispersion of electrons
in the sense that the momentum dependence of the Coulomb potential is strongly modified from the bare value
along that direction.
On the other hand, along the direction with quadratic dispersion of electrons,
the Coulomb potential is just weakly modified.
We call this the {\it anisotropic partial screening} phenomenon.
Strikingly, such a nontrivial screening effect makes the Coulomb interaction
{\it irrelevant} in the low energy limit.
Thus, the critical theory becomes a {\it non-interacting} fermion theory eventually,
which allows the complete theoretical understanding of the problem.
This behavior is in sharp contrast
to the cases of other SMs such as graphene and conventional Weyl SM with isotropic dispersion
where the Coulomb potential remains marginal in the infrared limit
rendering various physical quantities to receive
logarithmic temperature dependence.~\cite{Hosur,Isobe1,Isobe2,Chakravarty,Son}
The nontrivial screening effect of interacting AWF
can give rise to intriguing dielectric responses of the system,
which will be discussed below.

\section*{Model}
We begin with the description of the noninteracting AWF.
The effective Hamiltonian density has a simple form given by,
\begin{eqnarray}\label{eqn:H0}
H_0=-i\upsilon\partial_{1}\tau_{1}-i\upsilon\partial_{2}\tau_{2}-A\partial_{3}^{2}\tau_{3},
\end{eqnarray}
where $\upsilon$ is the velocity for the electron motion in $(k_{1},k_{2})$ plane
and $A$ is the inverse of the effective mass along the $k_{3}$ direction.
The Pauli matrices ($\tau_{1,2,3}$) are used to indicate the conduction and
valence bands.
Since the system lacks either the inversion symmetry or time reversal symmetry,
the gap-closing point is associated with only two bands in general.~\cite{Yang,Murakami1,Murakami2}
The topological phase transition from a Weyl SM to an insulator can be
described by adding the perturbation term $m \tau_3$. Depending on the relative sign between
$A$ and $m$, the ground state becomes either an insulator ($Am>0$) or
a Weyl SM ($Am<0$) with a pair of WPs having opposite chiralities.~\cite{Murakami1,Murakami2}
At the quantum critical point $(m=0)$,
the two WPs pair-annihilate at a certain momentum point
at which the net chirality should obviously be zero.
Such a zero chirality condition forces the electron dispersion
to be nonlinear at least along one direction, which leads
to the minimal Hamiltonian for the AWF as shown in (\ref{eqn:H0}).
In general, the quadratic dispersion occurs along the direction in which a pair of WPs migrate
before they are pair-annihilated.~\cite{Murakami1,Murakami2}
We provide a simple lattice model which captures all essential characteristics
of the topological phase transition between the Weyl SM and an insulator in Supplementary Note 1.

To describe the dynamics of AWF coupled to the long
range Coulomb interaction, we use the Euclidean path integral formalism and write
the effective action as follows.
\begin{eqnarray}
S=\int d^{4}x \Big[
\bar{\psi}\{\gamma_{0}(\partial_{0}+ig\varphi)+\upsilon(\gamma_{1}\partial_{1}+\gamma_{2}\partial_{2})
-A\partial_{3}^{2} \}\psi
+\frac{1}{2\sqrt{\eta}}\big\{(\partial_{1}\varphi)^{2}+(\partial_{2}\varphi)^{2}\big\}+\frac{\sqrt{\eta}}{2}(\partial_{3}\varphi)^{2}\Big].
\nonumber
\end{eqnarray}
As usual, the momentum cutoff ($\Lambda$) is implicitly assumed. The
matrices $\gamma_{0,1,2}$ , satisfying $\{\gamma_{i},\gamma_{j}\}=\delta_{ij}$,
are defined as
$\gamma_{0}=\tau_{3}$,$\gamma_{1}=\tau_{2}$,$\gamma_{2}=-\tau_{1}$. $\psi$
($\varphi$) represents an electron (boson) field operator and
$\bar{\psi}=\psi^{\dag}\gamma_{0}$. The Hubbard-Stratonovich transformation
introduces the bosonic field $\varphi$ for the instantaneous Coulomb
interaction. The dimensionless parameter $\eta$ is used to capture the
possible anisotropy of the Coulomb potential induced by the anisotropic fermion
dispersion. The electrons are coupled to the boson via the coupling constant
$g^{2}=e^{2}/\varepsilon$ where $e$ is the electric charge and $\varepsilon$ is
the dielectric constant.

Considering the anisotropic dispersion of electrons,
the coupling constants in the action naturally lead to three characteristic
energy scales, $E_{kin} = A^{-1} v^2$, $E_{c} =A^{-1} v g^2$, and
$E_{\Lambda} = \Lambda v$ which describe
the electron kinetic energy ($E_{kin}$) and Coulomb potential ($E_{c}$)
and the cutoff energy ($E_{\Lambda}$) related with the linear dispersion of electrons,
respectively.
The associated dimensionless coupling constants are
\begin{eqnarray}
\alpha = \frac{E_c}{E_{kin}}, \quad \beta
=\frac{\sqrt{\eta}}{12 \pi^2} \frac{E_c}{E_{\Lambda}}, \quad \gamma =\frac{
\alpha}{3\pi^{2}\sqrt{\eta}} \frac{E_{\Lambda}}{E_{kin}}.
\end{eqnarray}
Although only two of them are independent ($\gamma \sim \alpha^2/\beta$), for
notational convenience, we use the three coupling constants to perform
the renormalization group (RG) analysis.

The anisotropic fermion dispersion gives rise to unusual scaling properties
of the system. Let us first consider the noninteracting fermions with the
following scale transformation
\begin{eqnarray}\label{eqn:generalscaling}
&\tilde{t}=b^{-z}t,\quad \tilde{x}_{1,2}= b^{-z_{1}} x_{1,2},\quad
\tilde{x}_{3}= b^{-1}x_{3},
\end{eqnarray}
where the rescaled variable is indicated
by a tilde. Namely, the scaling dimension of $t$, $x_{1,2}$, and $x_{3}$ are
given by $[t]=-z$, $[x_{1,2}]=-z_{1}$, $[x_{3}]=-1$.
It is worth to note that
since the dispersion of the fermion is anisotropic while that of the boson is
isotropic, the scale invariance of the whole system requires to introduce the
general scaling dimension $z_{1}$. However, our RG results (below) are
independent of $z_{1}$ as it should be. The invariance of the fermionic part of
the action under the scale transformation in (\ref{eqn:generalscaling})
leads to the following scaling relations at the tree level, $[\psi] = z_{1}+1/2,
[v]=z-z_{1}, [A]=z-2$, $[g^{2}] = z - z_{1}, [\eta]=2z_{1}-2, [\varphi]={(z+z_{1})}/{2}$.
If we assume that $\upsilon$ is marginal while $A$
is either marginal or irrelevant, we obtain $z=z_{1}\leq2$, and $[g^{2}]=0$.
On the other hand, if $A$ is marginal while
$\upsilon$ is either marginal or irrelevant, $z=2$,
$z_{1}\geq2$, and $[g^{2}]\leq0$.
Therefore the Coulomb interaction is marginal for
$0<z_{1}\leq2$ while it is irrelevant otherwise,
hence we focus on the case of $0<z_{1}\leq2$ in the following.

\begin{figure*}[t]
\centering
\includegraphics[width=16 cm]{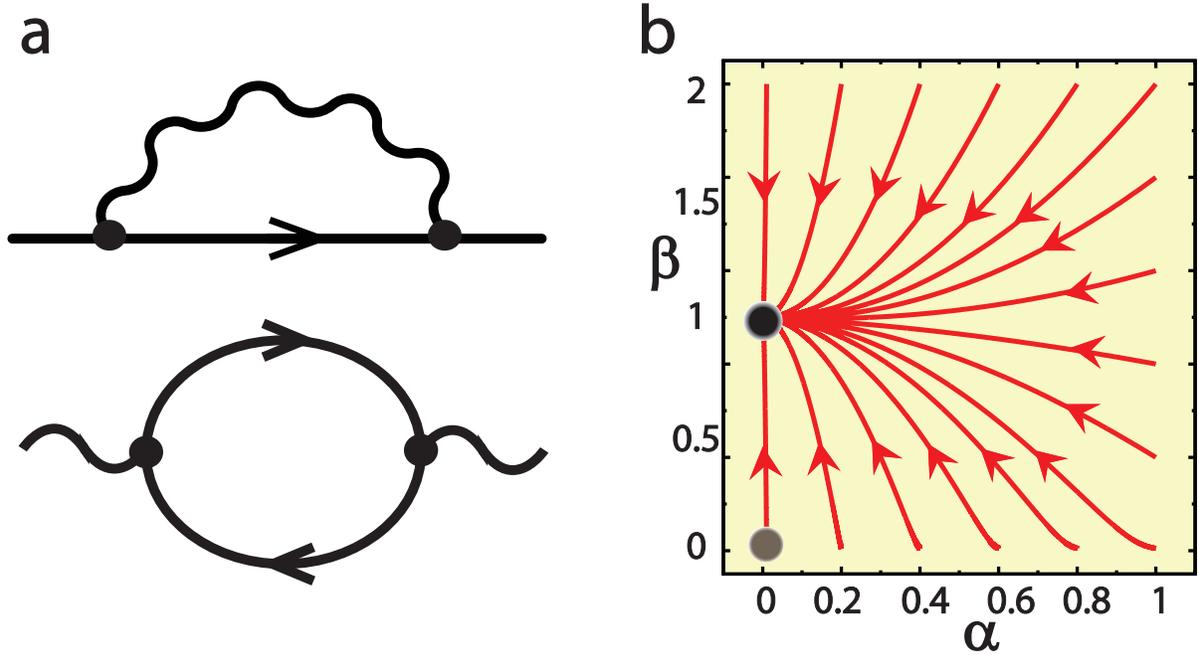}
\caption{
{\bf Feynman diagram for quantum corrections and
the resulting renormalization group flow.}
({\bf a}) Feynman diagrams for
electron self-energy (top) and boson self-energy (bottom).
The plain (wavy) line
is for electrons (bosons).
({\bf b}) The RG flow of the coupling constants
($\alpha$,$\beta$). The RG flow is obtained by solving the coupled
RG equations shown in the Supplementary Note 2.
The stable fixed point, marked by a black filled dot,
locates at $(\alpha,\beta)=(0,1)$, and the unstable one (the gray dot) locates
at $(0,0)$.
} \label{fig2}
\end{figure*}

{\it Renormalization group analysis.-}
We perform a momentum shell RG analysis of $S$ at one-loop
level. Because of the anisotropic dispersion, it is natural to consider two
different momentum cutoffs, i.e., $\Lambda_{\perp}$ for the dispersion in the
plane of $(k_{1}, k_{2})$ and $\Lambda_{3}$ for the $k_{3}$ dispersion. For
convenience, we set $\Lambda_{\perp}=\infty$ and perform the momentum-shell
integration in the following way,
$\int'\frac{d^{3}q}{(2\pi)^{3}}=\frac{1}{2\pi^{2}}\int_{\mu}^{\Lambda}dq_{3}\int_{0}^{\infty}q_{\perp}dq_{\perp}$
where $\Lambda=\Lambda_{3}$ and $\mu=\Lambda e^{-\ell}$.
In principle, three Feynmann diagrams need to be considered for the RG
calculation at one-loop order, i.e., the fermion self-energy, the boson
self-energy and the vertex correction. Since the fermion self-energy is
frequency independent for the instantaneous bare Coulomb interaction, the vertex
correction vanishes, consistent with the Ward identity in one-loop calculation.
Therefore, the two Feynman diagrams shown in Fig.~\ref{fig2}a determine the RG
flow.

The RG flows of the three coupling constants $\alpha$, $\beta$, $\gamma$ describe quantum
corrections from the interaction. Instead of showing the details, we only
present the beta functions leaving technical calculations to the Supplementary
Note 2.
\begin{eqnarray}\label{eqn:fullRG}
\frac{d\alpha}{d\ell}=\alpha\Big[-\frac{1}{16\pi}\alpha-\frac{1}{2}\beta-\frac{1}{2}\gamma \Big],\quad
\frac{d\beta}{d\ell}=\beta\Big[1-\frac{1}{8\pi}\alpha-\beta \Big],\quad
\frac{d\gamma}{d\ell}=\gamma\Big[-1-\gamma \Big].
\end{eqnarray}
Note that $z$ and $z_{1}$ are absent in (\ref{eqn:fullRG}) as expected.
The RG flow is illustrated in Fig.~\ref{fig2}b which shows two fixed points at
($\alpha,\beta $) = (0,0) and ($\alpha,\beta $) = (0,1), respectively. At both
fixed points, $\gamma=0$. The flow emerges from the former (unstable) fixed
point and terminates at the latter (stable) one. Interestingly, the
fine-structure constant ($\alpha$) is zero at both fixed points. The linear
$\alpha$ dependence of the electron self-energy,
\begin{eqnarray}
\Sigma_f
\sim\big[ v (k_1 \tau_1+k_2 \tau_2)+ 2A k_3^2 \tau_3 \big] (\frac{\alpha}{16\pi}
\ell) +O(\ell^2),
\end{eqnarray}
manifestly shows that the electrons basically
become free from interaction in the low energy limit.

The distinct natures of the two fixed points are clearly revealed in the
screening effect of the Coulomb potential. The inverse of the boson propagator
including the self-energy correction ($\Pi(\textbf{q})$) becomes,
\begin{eqnarray}
q_{\perp}^2+q_3^2-\Pi(\textbf{q})= q_{\perp}^{2}(1+\beta\ell)+q_3^2 (1+\gamma \ell)
+O(\ell^{2}). \nonumber
\end{eqnarray}
Here the unimportant $\eta$ dependence is
neglected. At the unstable fixed point, $(\alpha,\beta,\gamma)=(0,0,0)$, the Coulomb
potential decouples from electrons. On the other hand, at the stable fixed
point, $(\alpha,\beta,\gamma)=(0,1,0)$, the Coulomb potential receives a nontrivial
correction in the $q_{\perp}$ direction while there is no screening in the
$q_{3}$ direction. The non-trivial correction can be understood as the {\it
strong} screening effect with the anomalous dimension. Considering the different
scaling between the $q_{\perp}$ and $q_{3}$ directions, we obtain
\begin{eqnarray}
q_{\perp}^2+q_3^2-\Pi(\textbf{q}) \sim q_{\perp}^{2-1/z_{1}}+q_{3}^{2}, \nonumber
\end{eqnarray}
at the stable fixed point.
Thus, the Coulomb potential receives
{\it anisotropic partial screening} through the interaction with electrons.
Here $z_{1}$ can be fixed from the condition
that $v$ and $A$ do not run at the stable fixed point.
Then, the momentum anisotropy gives $z=z_1=2$ at the stable
fixed point. Therefore the renormalized Coulomb interaction has the following
structure $V_{C}(\textbf{q})\sim 1/(q_{\perp}^{3/2}+\eta q_{3}^{2})$ at the
fixed point. It means that in the real space
$V_{C}$ depends on the spatial coordinate $\textbf{r}=(\textbf{r}_{\perp},z)$ anisotropically as
$V_{C}(r_{\perp},z=0)\propto
r_{\perp}^{-5/4}$ and $V_{C}(r_{\perp}=0,z)\propto |z|^{-5/3}$.

\section*{Screening of a charged impurity}
\begin{figure*}[t]
\centering
\includegraphics[width=16 cm]{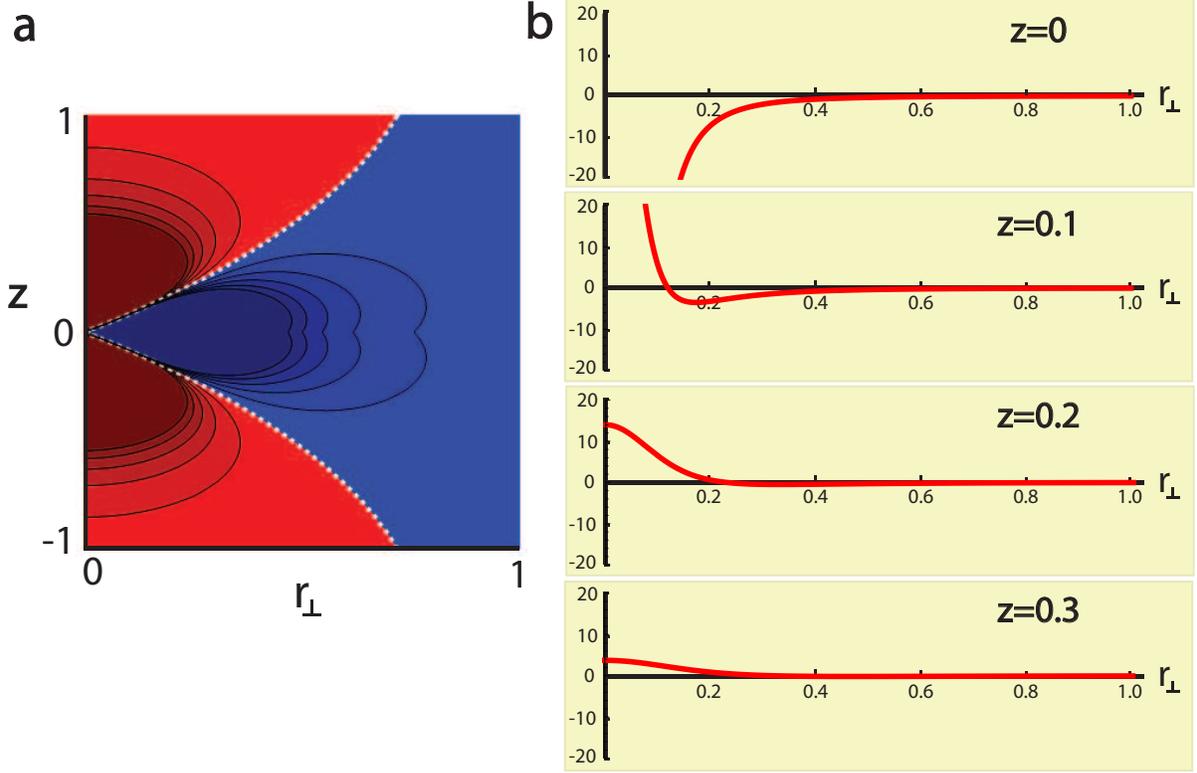}
\caption{
{\bf Distribution of the screening charge induced by a charged impurity in the noninteracting AWF.}
The distribution of the
induced charge $\rho_{\text{ind}}(\textbf{r})=\rho_{\text{ind}}(r_{\perp},z)$ in
real space when the charged impurity
is at the origin. Here $r_{\perp}$ and $z$ are measured in the unit of
$(B_{3}^{2}/B_{\perp}^{2})$ and it is assumed that
$g^{2}B_{\perp}^{6}/B_{3}^{5}=1$.
({\bf a})  The intensity plot of $\rho_{\text{ind}}$
in $(r_{\perp},z)$ plane. Here the white dotted line indicates the boundary on
which $\rho_{\text{ind}}=0$. $\rho_{\text{ind}}>0$ ($\rho_{\text{ind}}<0$) in
the red (blue) region on the left-hand (right-hand) side of the dotted line. The
color is darker when the magnitude of the charge density is higher.
({\bf b}) $r_{\perp}$ dependence of $\rho_{\text{ind}}$ when $z$ is fixed to 0, 0.1, 0.2,
and 0.3, respectively. This behavior will be totally changed when the Coulomb
interaction between electrons is considered, where the induced charge density is
strictly localized near the impurity site.
}\label{fig:inducedcharge}
\end{figure*}

The novel quantum effect in interacting AWF can induce
unusual dielectric responses of the system.
Here we consider the screening problem of a Coulomb impurity with
the electric charge $Ze$ ($e>0$) by AWF.
Because of the non-trivial quantum corrections caused by interacting fermions,
the distribution of the screening charge
in interacting AWF can be completely different from
that in the non-interacting system.
Let us first consider the
noninteracting AWF coupled to a single charged impurity, which can be
described by the following Hamiltonian,
\begin{eqnarray}
H=-i\upsilon\partial_{x}\sigma_{x}-i\upsilon\partial_{y}\sigma_{y}-A\partial_{z}^{2}\sigma_{z}-\frac{Zg_{0}^{2}}{4\pi
r},
\end{eqnarray}
where $g_{0}^{2}$ is the coupling constant of the noninteracting system.
In fact, because of the anisotropic momentum dependence of
the polarization function which can be summarized as
$\Pi(\textbf{q})=-B_{\perp}q_{\perp}^{3/2}-B_{3}q_{3}^{2}$ with the positive
constants $B_{\perp}$ and $B_{3}$,
the screening charges distribute in a highly unusual way even
in the noninteracting system.
According to the linear response theory, the induced charge density
is given by $\rho_{\text{ind}}(\textbf{q})=ZeV(\textbf{q})\Pi(\textbf{q})$
where $V(\textbf{q})=g_{0}^{2}/q^{2}$.~\cite{Kotov}
As shown in Fig.~\ref{fig:inducedcharge},
the peculiar structure of $\Pi(\textbf{q})$
gives rise to strong spatial anisotropy in the distribution of the induced charges in real space,
$\rho_{\text{ind}}(\textbf{r})\equiv\int \frac{d^{3}q}{(2\pi)^{3}}
e^{i\textbf{q}\cdot\textbf{r}}\rho_{\text{ind}}(\textbf{q})$.
In particular, owing to the anisotropy of $\rho_{\text{ind}}(\textbf{r})$,
the partially integrated charge
densities $Q_{z}(z)\equiv\int d^{2}r_{\perp}\rho_{\text{ind}}(\textbf{r})$ and
$Q_{\perp}(r_{\perp})\equiv\int dz \rho_{\text{ind}}(\textbf{r})$ show
characteristic coordinate dependence such as
\begin{eqnarray}
Q_{\perp}(r_{\perp})=-\frac{Zeg_{0}^{2}B_{\perp}}{4\pi^{2}}\frac{H(r_{\perp})}{r_{\perp}^{3/2}},\quad
Q_{z}(z)=-Zeg_{0}^{2}B_{3}\delta(z) \nonumber
\end{eqnarray}
where
$H(r_{\perp})\equiv2\pi\int_{0}^{x_{c}}dx\sqrt{x}J_{0}(x)$ in which $J_{0}(x)$
indicates the Bessel function of the first kind. Here $x_{c}$ is defined as
$x_{c}\equiv\Lambda_{\perp}r_{\perp}$ with the in-plane momentum cutoff
$\Lambda_{\perp}$. This sharp momentum cutoff causes the oscillation of
$H(r_{\perp})$ with fixed amplitude. Notice that the partially integrated charge
densities distribute in a highly unusual way, i.e., $Q_{z}(z)$ is strongly
localized near the impurity while $Q_{\perp}(r_{\perp})$ shows slow decay with a
power-law tail.
The detailed behavior of $\rho_{\text{ind}}(\textbf{r})$
is shown in Supplementary Note 3.

Surprisingly, the strong Coulomb interaction between electrons
completely modify the screening charge distribution in AWF.
Since all coupling constants are irrelevant at the stable fixed point, the
induced charge can be computed to the leading order
by using
$V_{\text{renorm}}(\textbf{q})=\frac{g^{2}}{q_{\perp}^{3/2}+\eta q_{3}^{2}}$ and
$\Pi(\textbf{q})=-B_{\perp}q_{\perp}^{3/2}-B_{3}q_{3}^{2}$.
Since the amount of the total
induced charge is fixed, i.e., $\int d^{2}r_{\perp}Q_{\perp}(r_{\perp})=\int
dzQ_{z}(z)$,
$B_{3}/B_{\perp}$ is fixed to be $\eta$, which gives rise to
$\rho_{\text{ind}}(\textbf{r})=-Zeg^{2}B_{\perp}\delta(\textbf{r})$.
In contrast to the case of the noninteracting system, the screening charge is
strictly localized near the impurity site. Similar localized distribution of the
induced charge is predicted in graphene.~\cite{Kotov_RMP} In graphene, it is
shown that the Coulomb interaction can induce small spreading of the screening
charge in subleading order. However, in our case such an additional spreading of
the screening charge can only play a minor role since the Coulomb interaction is
irrelevant at the fixed point
while it is marginal in graphene.~\cite{Son,Biswas}

\section*{Discussion}
Up to now we have considered the
system having a single AWF. However, in real materials,
several AWFs can simultaneously appear at the Fermi level due to the
time-reversal or crystalline symmetries. In this case, to describe the
renormalized Coulomb interaction, it is necessary to add the contribution from
all AWFs to the polarization. In other words, the polarization
function should be given by $\Pi_{\text{total}}=\sum_{i}\Pi_{i}$ where $\Pi_{i}$
indicates the polarization due to the $i$th AWF. Here
it is assumed that the polarization induced by the coupling between neighboring AWFs
can be neglected, which is valid if the distance between AWFs is far enough. The
important point is that the direction along which the AWF shows the quadratic
dispersion can vary for different AWFs in general. Therefore in the system with
multiple AWFs, the total polarization should behave as
$\sum_{i}\Pi_{i}(\textbf{q})\propto q^{3/2}$ on average with $q=|\textbf{q}|$.
In this case, the renormalized Coulomb interaction should have the following
form of $V_{C}(\textbf{q})\propto1/q^{3/2}$. Equivalently, it means that the
long range Coulomb interaction shows the unusual power law given by
$V_{C}(\textbf{r})\propto 1/|\textbf{r}|^{3/2}$ in real space due to the
screening effect. If we introduce a charged impurity to this system, the induced
charge spreads in a way of $\rho(r\gg1)\propto1/r^{2}$ when the
electron-electron interaction is neglected. On the other hand, once the
electron-electron interaction is considered, the induced charge is again
strictly localized near the impurity site, i.e., $\rho(r)=Z'e\delta(r)$ with a
constant $Z'$.

The results from our perturbative RG approach are further
supported by the random phase approximation (RPA) approach.
As shown in the Supplementary Note 4, we find that
along the direction with linear (quadratic) dispersion,
the momentum dependence of the polarization shows
$\Pi(q_{\perp},0)\propto q_{\perp}^{3/2}$ ($\Pi(0,q_{3})\propto
q_{3}^{2}$).
Note that the RPA and the one-loop RG calculations match perfectly,
thus we believe that the results of the one-loop RG calculation is {\it robust}.
Besides the fact that two different methods give the same momentum dependence
of $\Pi(\textbf{q})$,
we further consider the corrections to the RPA results by performing
the strong coupling expansion.
As shown in Supplementary Note 5, it is explicitly proved that the quantum
fluctuations near the stable fixed point do not cause any divergent corrections
confirming the robustness of the conclusions obtained from the RG analysis.

To conclude, we have rigorously demonstrated the novel quantum
critical phenomena of the AWF.
We emphasize that the anisotropic partial screening is clearly distinct from the
conventional screening phenomena in usual metallic or semi-metallic systems,
which is concisely summarized in Table 1.
For example, the Thomas Fermi screening in conventional metals
forces the renormalized Coulomb interaction to be short-ranged
and the fermionic excitations coupled by the short range repulsive interaction
gives rise to the Fermi liquid state.~\cite{Shankar}
On the other hand,
in the AWF, the interplay between the Coulomb interaction and fermions
induces the anisotropic partial screening, which allows
the screened Coulomb interaction to maintain the long-ranged nature and
the fermions to be free from the Coulomb interaction in the low energy limit.
Therefore the anisotropic partial screening can be considered as
the intermediate screening phenomena
distinguished from both the Thomas-Fermi screening in metals and
the logarithmic screening in Dirac fermions,
which is uncovered for the first time in this study.

%\newpage
\bibliographystyle{naturemag}

%\bibliography{References}

{\small \subsection*{Acknowledgements}
We are grateful for support from the Japan Society for the Promotion of Science (JSPS) through the `Funding Program for World-Leading Innovative R\&D on Science and Technology (FIRST Program)}, and Grant-in-Aids for Scientific Research (No. 24224009) from the
Ministry of Education, Culture, Sports, Science and Technology (MEXT).
B.-J. Yang and N. Nagaosa greatly appreciate the stimulating discussions with Ammon Aharony, Ora Entin-Wohlman,
Michael Hermele, and Miguel A. Cazalilla.
E.-G. Moon is grateful for invaluable discussion
with Leon Balnets, Max Metlitski, and Cenke Xu
and supported by the MRSEC Program of the National Science Foundation under Award No. DMR 1121053.

%{\small \subsection*{Author contributions}
%N. N. conceived the original ideas.
%B.-J.Y., E. -G. M., H. I. did the calculations.
%All authors analyzed
%the data and wrote the manuscript.
%}

%{\small \subsection*{Additional information}
%Supplementary Information is available in the {\bf online version of the paper.}
%Reprints and permissions information is available online at {\bf http://www.nature.com/reprints.}
%Correspondence and requests for materials should be addressed to B.-J.Y. or N.N..
%}

%{\small \subsection*{Competing financial interests}
%The authors declare no competing financial interests.
%}

\
\newpage
%%%%%%%%%%%%%%%%%%%%%%%%%%%%%%%%%%%%%%%%%%%%%%%%%%%%%%%%%%%%%%%%%%%
\begin{table*}[h]
\begin{tabular}{@{}|c|c|c|c|}
\hline
& \quad $V_{\text{FP}}(\textbf{q})$ \quad & $\alpha_{\text{FP}}$ & Fermionic Excitations \\
\hline
\hline
Metal & $\frac{1}{q^{2}+q_{\text{TF}}^{2}}$  & Marginal & Fermi Liquid\\
\hline
Anisotropic Weyl fermion (AWF) & $\frac{1}{q_{\perp}^{3/2}+q_{3}^{2}}$  & Irrelevant & Fermi Liquid\\
\hline
Weyl SM & $\frac{1}{q^{2}}$  & Marginally Irrelevant & (Marginal) Fermi Liquid\\
\hline
\end{tabular}
\end{table*}
{\noindent {\bf Table 1:}
{\bf Physical properties of interacting metallic or semi-metallic
systems in 3D coupled with the instantaneous Coulomb interaction.}
Here ``Metal" indicates the conventional metal with Fermi surface and
``Weyl SM" means the 3D semi-metal with linear dispersion
in all spatial directions.
$V_{\text{FP}}(\textbf{q})$ ($\alpha_{\text{FP}}$) indicates
the screened Coulomb potential (renormalized coupling constant)
at the stable fixed point.
As for the $\textbf{q}$ dependence of $V_{\text{FP}}(\textbf{q})$,
$q^{2}=q_{1}^{2}+q_{2}^{2}+q_{3}^{2}=q_{\perp}^{2}+q_{3}^{2}$
and $1/q_{\text{TF}}$ indicates the Thomas Fermi screening length.
When the electron dispersion is quadratic in all directions,
it is shown that the screened Coulomb
potential becomes $V(\textbf{q})\sim 1/q$ and a non-Fermi liquid ground state
can be realized. (Ref. [\onlinecite{Moon}])
}
%%%%%%%%%%%%%%%%%%%%%%%%%%%%%%%%%%%%%%%%%%%%%%%%%%%%%%%%%%%%%%%%%%%
\
\newpage
%%%%%%%%%%%%%%%%%%%%%%%%%%%%%%%%%%%%%%%%%%%%%%%%%%%%%%%%%%%%%%%%%%%
%%%%%%%%%%%%%%%%%%%%%%%%%%%%%%%%%%%%%%%%%%%%%%%%%%%%%%%%%%%%%%%%%%%
\
\newpage
%%%%%%%%%%%%%%%%%%%%%%%%%%%%%%%%%%%%%%%%%%%%%%%%%%%%%%%%%%%%%%%%%%%
%%%%%%%%%%%%%%%%%%%%%%%%%%%%%%%%%%%%%%%%%%%%%%%%%%%%%%%%%%%%%%%%%%%
\
\newpage
%%%%%%%%%%%%%%%%%%%%%%%%%%%%%%%%%%%%%%%%%%%%%%%%%%%%%%%%%%%%%%%%%%%
%%%%%%%%%%%%%%%%%%%%%%%%%%%%%%%%%%%%%%%%%%%%%%%%%%%%%%%%%%%%%%%%%%%%
\
\newpage
\noindent{\bf Supplementary Note 1}
\\
{\bf  Lattice model for anisotropic Weyl fermion.}

Here we provide a simple lattice model Hamiltonian
describing the topological phase transition from a Weyl SM
to an insulator. (See also Ref.[26].) The AWF appears
at the quantum critical point. We consider spinless fermions on a tetragonal lattice
with two orbital degrees of freedom in the unit cell, which can be
represented by
\begin{eqnarray}
H=2t_{x}\sin k_{x}\sigma_{x} + 2t_{y}\sin k_{y}\sigma_{y}
+[2-\cos k_{x}-\cos k_{y}+\frac{1}{2}\cos k_{z}-m]\sigma_{z},
\end{eqnarray}
where $\sigma_{x,y,z}$ are the Pauli matrices
indicating two orbital degrees of freedom and $t_{x}$, $t_{y}$ are the hopping integrals
along the $x$, $y$ directions,
respectively. Here we assume $|m|<3/2$ for convenience.
In this system, when $|m|<1/2$, the Weyl SM appears as the ground state
which has a pair of Weyl points with opposite
chirality along the $k_{z}$ axis.
The critical point at $m_{c1}=-1/2$ ($m_{c2}=1/2$) mediates
the quantum phase transition from the Weyl SM
to a trivial band insulator (a 3D quantum Hall insulator)
which exists for $-3/2<m<-1/2$ ($1/2<m<3/2$).
At these two quantum critical points,
the bulk energy spectrum shows the quadratic dispersion
along the $k_{z}$ direction while it is linear
along the $(k_{x},k_{y})$ plane,
which indicates the emergence of the AWF.

\newpage
\noindent{\bf Supplementary Note 2}
\\
{\bf  Renormalization Group analysis.}

In this section, we provide a detailed description of the RG analysis. To
perform the one-loop renormalization group analysis, we write the Euclidean
effective action describing the low energy electrons coupled to long range
Coulomb interaction in the following way,
\begin{eqnarray}
S_{\text{bare}}=\int
d^{4}x \Big[
\bar{\psi}\{\gamma_{0}(\partial_{0}+ig\varphi)+\upsilon(\gamma_{1}\partial_{1}+\gamma_{2}\partial_{2})
-A\partial_{3}^{2} \}\psi
+\frac{1}{2\sqrt{\eta}}\big\{(\partial_{1}\varphi)^{2}+(\partial_{2}\varphi)^{2}\big\}+\frac{\sqrt{\eta}}{2}(\partial_{3}\varphi)^{2}\Big],
\nonumber
\end{eqnarray}
which is basically the same as the action in the main text of the
paper. Here the instantaneous Coulomb interaction term is decoupled by
introducing the bosonic Hubbard-Stratonovich field $\varphi$. The
non-interacting electron and boson Green's functions are given by
\begin{eqnarray}
G_{0}(k_{0},\textbf{k})&=&\frac{-i[\gamma_{0}k_{0}+\upsilon(\gamma_{1}k_{1}+\gamma_{2}k_{2})]+Ak_{3}^{2}}
{k_{0}^{2}+\upsilon^{2}k_{\perp}^{2}+A^{2}k_{3}^{4}}, \nonumber\\
D_{0}(k_{0},\textbf{k})&=&\frac{\sqrt{\eta}}{k_{\perp}^{2}+ \eta k_{3}^{2}},\nonumber
\end{eqnarray}
where $k_{\perp}^{2}=k_{1}^{2}+k_{2}^{2}$.

{\bf Boson selfenergy.-}
At first we compute the one-loop boson self-energy
$\Pi(k)$, which is given by
\begin{eqnarray}\label{eqn:spl_polarization}
\Pi(k)&=&g^{2}\int_{-\infty}^{\infty}\frac{dq_{0}}{2\pi}\int'\frac{d^{3}q}{(2\pi)^{3}}
\text{Tr}[\gamma_{0}G_{0}(q)\gamma_{0}G_{0}(k+q)],\nonumber \\
&=&-2g^{2}\int_{-\infty}^{\infty}\frac{dq_{0}}{2\pi}\int'\frac{d^{3}q}{(2\pi)^{3}}
\frac{[q_{0}(q_{0}+k_{0})-\upsilon^{2}\textbf{q}_{\perp}\cdot(\textbf{q}_{\perp}+\textbf{k}_{\perp})-A^{2}q_{3}^{2}(q_{3}+k_{3})^{2}]}
{[q_{0}^{2}+E_{q}^{2}][(q_{0}+k_{0})^{2}+E_{q+k}^{2}]},
\end{eqnarray}
where
$E_{k}=\sqrt{\upsilon^{2}k_{\perp}^{2}+A^{2}k_{3}^{4}}$. Here
$\int'\frac{d^{3}q}{(2\pi)^{3}}$ indicates the integral over the momentum shell.
To determine the renormalization of the bare boson propagator $D_{0}(k)$ by
$\Pi(k)$, we expand $\Pi(k)$ in powers of $k_{i}$ up to quadratic order, which
leads to
\begin{eqnarray}\label{eqn:spl_polarization_smallk}
\Pi(k)=-\frac{g^{2}}{8}(\upsilon^{2}k_{\perp}^{2})\int'\frac{d^{3}q}{(2\pi)^{3}}\Big[\frac{2}{E_{q}^{3}}
-\frac{\upsilon^{2}q_{\perp}^{2}}{E_{q}^{5}}\Big] -g^{2}\upsilon^{2}(A^{2}k_{3}^{2})\int'\frac{d^{3}q}{(2\pi)^{3}}
\frac{q_{3}^{2}q_{\perp}^{2}}{E_{q}^{5}}.
\end{eqnarray}
After a
straightforward calculation of the momentum-shell integral, we can obtain
\begin{eqnarray}\label{eqn:spl_photon_selfenergy}
\Pi(k)=-k_{\perp}^{2}\ell
\frac{1}{\sqrt{\eta}}I_{4}-k_{3}^{2}\ell \sqrt{\eta}I_{5},
\end{eqnarray}
in which
\begin{eqnarray}
I_{4}=\frac{1}{12\pi^{2}}\frac{g^{2}}{\upsilon}\Big(\frac{\upsilon\sqrt{\eta}}{A\Lambda}\Big),\quad
I_{5}=\frac{1}{3\pi^{2}}\frac{g^{2}}{\upsilon}\Big(\frac{A\Lambda}{\upsilon\sqrt{\eta}}\Big).
\end{eqnarray}

{\bf Electron selfenergy.-}
Now let us consider the renormalization of
$\upsilon$ and $A$ by computing the electron self-energy $\Sigma(k)$. At
one-loop order, $\Sigma(k)$ can be written as
\begin{eqnarray}
\Sigma(k_{0},\textbf{k}_{\perp},k_{3}) &=&-g^{2}\int_{-\infty}^{\infty}\frac{dq_{0}}{2\pi}\int^{'}\frac{d^{3}q}{(2\pi)^{3}}
\gamma_{0}G_{0}(q)\gamma_{0}D_{0}(k-q) \nonumber\\
&=&-\frac{g^{2}\sqrt{\eta}}{2}\int^{'}\frac{d^{3}q}{(2\pi)^{3}}
\frac{[i\upsilon(\gamma_{1}q_{1}+\gamma_{2}q_{2})+Aq_{3}^{2}]}{E_{q}} \frac{1}{[(\textbf{k}_{\perp}-\textbf{q}_{\perp})^{2}+\eta(k_{3}-q_{3})^{2}]}, \nonumber
\end{eqnarray}
which, for small $\textbf{k}_{\perp}$ and $k_{3}$, can be written as
\begin{eqnarray}
\Sigma(\textbf{k}_{\perp},k_{3})
=-\frac{g^{2}\sqrt{\eta}}{2}i\upsilon(\gamma_{1}k_{1}+\gamma_{2}k_{2})J_{1}
-\frac{g^{2}}{2}\eta^{3/2} k_{3}^{2}(J_{2}+J_{3}).\nonumber
\end{eqnarray}
Here $J_{1,2,3}$ are as follows,
\begin{eqnarray}
J_{1}&\equiv&\int^{'}\frac{d^{3}q}{(2\pi)^{3}}
\frac{1}{\sqrt{\upsilon^{2}q^{2}_{\perp}+A^{2}q_{3}^{4}}}
\frac{q_{\perp}^{2}}{(q_{\perp}^{2}+\eta q_{3}^{2})^{2}} \nonumber\\
&=&\frac{1}{4\pi^{2}}\int_{\mu}^{\Lambda}dq_{3}\int_{0}^{\infty}d(q_{\perp}^{2})
\frac{1}{\sqrt{\upsilon^{2}q^{2}_{\perp}+A^{2}q_{3}^{4}}}
\frac{q_{\perp}^{2}}{(q_{\perp}^{2}+\eta q_{3}^{2})^{2}} \nonumber\\
&=&\frac{\ell}{4\pi^{2}\upsilon\eta^{1/2}}
\Big[\frac{A\Lambda}{\upsilon\eta^{1/2}}\frac{1}{1-\bar{A}^{2}}
+\cos^{-1}\Big(\bar{A}\Big) \frac{(1-2\bar{A}^{2})}{(1-\bar{A}^{2})^{3/2}}
\Big],\nonumber
\end{eqnarray}
where $\ell=\ln\frac{\Lambda}{\mu}$ is the usual RG scale
parameter and $\bar{A}\equiv\frac{A\Lambda}{\upsilon\sqrt{\eta}}$. If we solve
the full coupled RG equation, it can be shown that
$\bar{A}^{2}=\frac{A^{2}\Lambda^{2}}{\upsilon^{2}\eta}\ll 1$. Therefore we can
make an expansion in powers of $\bar{A}$ and the leading contribution to $J_{1}$
can be written as
\begin{eqnarray}
J_{1}\approx\frac{1}{8\pi\upsilon}\frac{1}{\sqrt{\eta}}\ln\frac{\Lambda}{\mu}.
\end{eqnarray}
The other integrals appearing in $\Sigma(k)$ can be done in the
following way,
\begin{eqnarray}
J_{2}&\equiv&\int^{'}\frac{d^{3}q}{(2\pi)^{3}}
\frac{Aq_{3}^{2}}{\sqrt{\upsilon^{2}q^{2}_{\perp}+A^{2}q_{3}^{4}}}
\Big[-\frac{1}{(q_{\perp}^{2}+\eta q_{3}^{2})^{2}}\Big] \nonumber\\
&=&-\frac{A\ell}{4\pi^{2}\upsilon\eta^{3/2}}
\Big[\frac{\bar{A}}{\bar{A}^{2}-1} +\cos^{-1}\Big(\bar{A}\Big)
\frac{1}{(1-\bar{A}^{2})^{3/2}} \Big] \approx
-\frac{A}{8\pi\upsilon}\frac{1}{\eta^{3/2}}\ln\frac{\Lambda}{\mu}.
\end{eqnarray}
Similarly,
\begin{eqnarray}
J_{3}&\equiv&\int^{'}\frac{d^{3}q}{(2\pi)^{3}}
\frac{Aq_{3}^{2}}{\sqrt{\upsilon^{2}q^{2}_{\perp}+A^{2}q_{3}^{4}}}
\Big[\frac{4\eta q_{3}^{2}}{(q_{\perp}^{2}+\eta q_{3}^{2})^{3}}\Big]
\nonumber\\
&=&\frac{A\ell}{4\pi^{2}\upsilon\eta^{3/2}}
\Big[\frac{A\Lambda}{\upsilon\eta^{1/2}}\frac{(2\bar{A}^{2}-5)}{(\bar{A}^{2}-1)^{2}}
+3\cos^{-1}(\bar{A}) \frac{1}{(1-\bar{A}^{2})^{5/2}} \Big]
\approx\frac{3A}{8\pi\upsilon}\frac{1}{\eta^{3/2}}\ln\frac{\Lambda}{\mu}.\nonumber
\end{eqnarray}
Then $\Sigma(k)$ can be written as
\begin{eqnarray}\label{eqn:spl_electron_selfenergy}
\Sigma(k)=-i\upsilon(\gamma_{1}k_{1}+\gamma_{2}k_{2})\ell I_{1} -Ak_{3}^{2}\ell
I_{3}
\end{eqnarray}
$I_{1}$ ($I_{3}$) describes the renormalization of $\upsilon$
($A$). Explicitly, for small $\bar{A}$ limit, $I_{1,3}$ are given by
\begin{eqnarray}
I_{1}=\frac{1}{16\pi}\frac{g^{2}}{\upsilon},\quad \quad
I_{3}=\frac{1}{8\pi}\frac{g^{2}}{\upsilon}.
\end{eqnarray}

{\bf RG equations.-} Including $\Sigma(k)$ and $\Pi(k)$, the full effective
action can be written as
\begin{eqnarray}\label{eqn:spl_renormalized_action}
S_{\text{renorm}}&=&S_{\text{bare}}+\int d^{4}x \bar{\psi}(-\Sigma)\psi + \int
d^{4}x \frac{1}{2}\varphi(-\Pi)\varphi \nonumber\\
&=&\int d^{4}x \Big[
\bar{\psi}\{\gamma_{0}(\partial_{0}+ig\varphi) +\upsilon[1+\ell
I_{1}](\gamma_{1}\partial_{1}+\gamma_{2}\partial_{2}) -A[1+\ell
I_{3}]\partial_{3}^{2} \}\psi \nonumber\\ &&\qquad+\frac{1}{2\sqrt{\eta}}[1+\ell
I_{4}][(\partial_{1}\varphi)^{2}+(\partial_{2}\varphi)^{2}]
+\frac{1}{2}\sqrt{\eta}[1+\ell I_{5}](\partial_{3}\varphi)^{2}\Big],
\end{eqnarray}

The one loop RG equations can be obtained in the following way. At first, we
rescale the space-time coordinates,
\begin{eqnarray}
x_{0}=\tilde{x}_{0}b^{z},
\quad x_{1}=\tilde{x}_{1}b^{z_{1}}, \quad x_{2}=\tilde{x}_{2}b^{z_{1}}, \quad
x_{3}=\tilde{x}_{3}b,
\end{eqnarray}
where the tilde is used to indicate the scaled
variable. Then we introduce renormalization constants $Z_{\psi}$, $Z_{\varphi}$,
$Z_{g}$, $Z_{\upsilon}$, $Z_{\bar{A}}$, and $Z_{\eta}$ in the following way,
\begin{eqnarray}
\psi &=&Z_{\psi}^{-1/2}\tilde{\psi},\quad\varphi=Z_{\varphi}^{-1/2}\tilde{\varphi},\quad
g=Z_{g}^{-1/2}\tilde{g}, \nonumber\\
\upsilon &=& Z_{\upsilon}^{-1}\tilde{\upsilon},\quad
\bar{A}=Z_{\bar{A}}^{-1}\tilde{\bar{A}},\quad\eta=Z_{\eta}^{-1}\tilde{\eta},
\end{eqnarray}
Then we assume that the renormalized action $S_{\text{renorm}}$ has
the same form as the bare action $S_{\text{bare}}$, which leads to
\begin{eqnarray}
\frac{d\ln\upsilon}{d\ell}&=&\frac{1}{\upsilon}\frac{d\upsilon}{d\ell}=z-z_{1}+I_{1},
\nonumber\\
\frac{d\ln A}{d\ell}&=&z-2+I_{3}, \nonumber\\
\frac{d\ln g^{2}}{d\ell}&=&z-z_{1}-\frac{1}{2}I_{4}-\frac{1}{2}I_{5},
\nonumber\\
\frac{d\ln\eta}{d\ell}&=&2z_{1}-2-I_{4}+I_{5}.\nonumber
\end{eqnarray}
At this point, let us
explain the physical meaning of dimensionless variables
$\alpha\equiv\frac{g^{2}}{\upsilon}$ and $\bar{A}\equiv
\frac{A\Lambda}{\upsilon\sqrt{\eta}}$. Here $\alpha$ indicates the strength of
the Coulomb interaction relative to the kinetic energy originating from the
linear dispersion. $\bar{A}$ compares the kinetic energy from the quadratic
dispersion relative to the kinetic energy coming from the linear dispersion. In
terms of $\alpha$ and $\bar{A}$, $I_{1,3,4,5}$ can be written in the following
way.
\begin{eqnarray}\label{eqn:I}
I_{1}&=&\frac{\alpha}{8\pi^{2}}\Big[\frac{\bar{A}}{1-\bar{A}^{2}}+\cos^{-1}(\bar{A})\frac{(1-2\bar{A}^{2})}{(1-\bar{A}^{2})^{3/2}}\Big]
\approx\frac{1}{16\pi}\alpha, \nonumber\\
I_{3}&=&\frac{\alpha}{8\pi^{2}}\Big[\frac{\bar{A}(\bar{A}^{2}-4)}{(\bar{A}^{2}-1)^{2}}\
+\cos^{-1}(\bar{A})\frac{(2+\bar{A}^{2})}{(1-\bar{A}^{2})^{5/2}}\Big]
\approx\frac{1}{8\pi}\alpha, \nonumber\\
I_{4}&=&\frac{1}{12\pi^{2}}\frac{g^{2}}{\upsilon}\Big(\frac{\upsilon\sqrt{\eta}}{A\Lambda}\Big)=\frac{1}{12\pi^{2}}\frac{\alpha}{\bar{A}},
\nonumber\\
I_{5}&=&\frac{1}{3\pi^{2}}\frac{g^{2}}{\upsilon}\Big(\frac{A\Lambda}{\upsilon\sqrt{\eta}}\Big)=\frac{1}{3\pi^{2}}\alpha\bar{A},
\end{eqnarray}
In the small $\bar{A}$ limit the above RG equations reduce to
\begin{eqnarray}\label{eqn:spl_RGsmallA}
\frac{d\ln\upsilon}{d\ell}&=&z-z_{1}+\frac{\alpha}{16\pi}, \nonumber\\
\frac{d\ln A}{d\ell}&=&z-2+\frac{\alpha}{8\pi}, \nonumber\\
\frac{d\ln g^{2}}{d\ell}&=&z-z_{1}-\frac{1}{24\pi^{2}}\frac{\alpha}{\bar{A}}-\frac{1}{6\pi^{2}}\alpha\bar{A},
\nonumber\\
\frac{d\ln\eta}{d\ell}&=&2z_{1}-2-\frac{1}{12\pi^{2}}\frac{\alpha}{\bar{A}}+\frac{1}{3\pi^{2}}\alpha\bar{A}.
\end{eqnarray}

Similarly, the RG equations for $\alpha$ and $\bar{A}$ can be written as
\begin{eqnarray}\label{eqn:spl_fullRG}
\frac{d\ln\alpha}{d\ell}&=&-I_{1}-\frac{1}{2}I_{4}-\frac{1}{2}I_{5} \nonumber\\
&=&\Big[-\frac{1}{16\pi}\alpha-\frac{1}{24\pi^{2}}\frac{\alpha}{\bar{A}}-\frac{1}{6\pi^{2}}\alpha\bar{A}\Big],
\nonumber\\
\frac{d\ln\bar{A}}{d\ell}&=&-1-I_{1}+I_{3}+\frac{1}{2}I_{4}-\frac{1}{2}I_{5}
\nonumber\\
&=&-1+\Big[\frac{1}{16\pi}\alpha+\frac{1}{24\pi^{2}}\frac{\alpha}{\bar{A}}-\frac{1}{6\pi^{2}}\alpha\bar{A}\Big].
\end{eqnarray}
The RG equations can be neatly expressed by using $\alpha$, $\beta$,
$\gamma$ which are defined as
\begin{eqnarray}
\alpha =\frac{E_c}{E_{kin}},\quad \beta =\frac{\sqrt{\eta}}{12
\pi^2}\frac{E_c}{E_{\Lambda}},\quad \gamma
=\frac{\alpha}{3\pi^{2}\sqrt{\eta}}\frac{E_{\Lambda}}{E_{kin}},
\end{eqnarray}
where $E_{kin} = A^{-1} \upsilon^2 \, , \, E_{c} =A^{-1} \upsilon g^2 \, , \,
E_{\Lambda} = \Lambda \upsilon$. Here $\beta$ and $\gamma$ can be rewritten in
terms of $\alpha$ and $\bar{A}$ as follows,
\begin{eqnarray}
\beta
=\frac{1}{12\pi^{2}}\frac{\alpha}{\bar{A}},\quad \gamma
=\frac{1}{3\pi^{2}}\alpha\bar{A}.
\end{eqnarray}
The RG equations for $\alpha$,
$\beta$, $\gamma$ are shown in the main text.

\newpage
\noindent{\bf Supplementary Note 3}
\\
{\bf The distribution of the induced charge.}

Here we show the detailed structure of the induced charge
$\rho_{\text{ind}}(\textbf{r})=\int \frac{d^{3}q}{(2\pi)^{3}}e^{i\textbf{q}\cdot
\textbf{r}}\rho_{\text{ind}}(\textbf{q})$. After a straightforward computation
of the momentum integration, we can obtain the following result.
\begin{eqnarray}
\rho_{\text{ind}}(\textbf{r})=\int \frac{d^{3}q}{(2\pi)^{3}}e^{i\textbf{q}\cdot
\textbf{r}}\rho_{\text{ind}}(\textbf{q})
=\rho_{\text{ind}}^{(\text{I})}(\textbf{r})+\rho_{\text{ind}}^{(\text{II})}(\textbf{r}),
\end{eqnarray}
in which
\begin{eqnarray}
\rho_{\text{ind}}^{(\text{I})}(\textbf{r})&=&
-\frac{Zg^{2}B_{\perp}}{8\pi^{3/2}}\frac{|z|}{r^{7/2}}
\big\{8\emph{E}(\frac{1}{2}-\frac{|z|}{2r})-
(4+\frac{r}{|z|})\emph{K}(\frac{1}{2}-\frac{|z|}{2r}) \big\}, \nonumber\\
\rho_{\text{ind}}^{(\text{II})}(\textbf{r})&=&
\frac{Zg^{2}B_{3}}{4\pi}\frac{(2z^{2}-r_{\perp}^{2})}{r^{5}}, \nonumber
\end{eqnarray}
where
$r=\sqrt{r_{\perp}^{2}+z^{2}}$. Here $\emph{K}$ ($\emph{E}$) indicates the
complete elliptic integral of the 1st (2nd) kind.

When $r_{\perp}=0$, we can obtain that
\begin{eqnarray}
\rho_{\text{ind}}(r_{\perp}=0,z)=-\frac{3Zg^{2}B_{\perp}}{16\sqrt{\pi}}
\frac{1}{|z|^{5/2}}+\frac{Zg^{2}B_{3}}{2\pi}\frac{1}{|z|^{3}}. \nonumber
\end{eqnarray}
On the other hand, when $z=0$,
\begin{eqnarray}
\rho_{\text{ind}}(r_{\perp},z=0)=\frac{Zg^{2}B_{\perp}\emph{K}(1/2)}{8\pi^{3/2}}
\frac{1}{r_{\perp}^{5/2}}-\frac{Zg^{2}B_{3}}{4\pi}\frac{1}{r_{\perp}^{3}}.\nonumber
\end{eqnarray}
Therefore we can clearly see that
$\rho_{\text{ind}}(r_{\perp}\rightarrow0,z=0)$ and
$\rho_{\text{ind}}(r_{\perp}=0,z\rightarrow0)$ have the opposite sign, which is
consistent with the charge distribution shown in Fig. 3a.

\newpage
\noindent{\bf Supplementary Note 4}
\\
{\bf Screened Coulomb interaction in RPA analysis.}

To confirm the anomalous structure of the effective Coulomb interaction at the
stable fixed point, here we compute the polarization function $\Pi(k)$ in the
following way.
\begin{eqnarray}
\Pi(k)&=&g^{2}\int_{-\infty}^{\infty}\frac{dq_{0}}{2\pi}\int\frac{d^{3}q}{(2\pi)^{3}}
\text{Tr}[\gamma_{0}G_{0}(q)\gamma_{0}G_{0}(k+q)], \nonumber\\
&=&-2g^{2}\int_{-\infty}^{\infty}\frac{dq_{0}}{2\pi}\int\frac{d^{3}q}{(2\pi)^{3}}
\frac{[q_{0}(q_{0}+k_{0})-\upsilon^{2}\textbf{q}_{\perp}\cdot(\textbf{q}_{\perp}+\textbf{k}_{\perp})-A^{2}q_{3}^{2}(q_{3}+k_{3})^{2}]}
{[q_{0}^{2}+E_{q}^{2}][(q_{0}+k_{0})^{2}+E_{q+k}^{2}]}\nonumber
\end{eqnarray}
In contrast to the case of the RG approach where the integration is
performed on the momentum shell, we perform the momentum integration over the
full range of $\textbf{q}$ and extract the $k$ dependence of $\Pi(k)$.

After the integration over $q_{0,1,2}$, the static polarization
$\Pi(k_{0}=0,\vec{k})$ can be written as
\begin{eqnarray}
\Pi(k_{0}=0,\frac{\vec{k}_{\perp}}{\upsilon},\frac{k_{3}}{\sqrt{A}})
&=&-\frac{g^{2}}{8\pi^{2}\upsilon^{2}\sqrt{A}}\int^{1}_{0}dx\int^{\infty}_{-\infty}dq_{3}
\nonumber\\
&\times&\frac{[2x(1-x)k_{\perp}^{2}-q_{3+}^{2}q_{3-}^{2}+xq_{3+}^{4}+(1-x)q_{3-}^{4}]}{\sqrt{x(1-x)k_{\perp}^{2}+xq_{3+}^{4}+(1-x)q_{3-}^{4}}}
\nonumber
\end{eqnarray}
where $q_{3\pm}=q_{3}\pm\frac{1}{2}k_{3}$. To perform the
integration over $q_{0,1,2}$, we have used the following formula of dimensional
regularization
\begin{eqnarray}
\int\frac{d^{d}k}{(2\pi)^{d}}\frac{1}{(k^{2}+\Delta)^{n}}
&=&\frac{1}{(4\pi)^{d/2}}\frac{\Gamma(n-d/2)}{\Gamma(n)}\frac{1}{\Delta^{n-d/2}},
\nonumber\\ \int \frac{d^{d}k}{(2\pi)^{d}}\frac{k^{2}}{(k^{2}+\Delta)^{n}}
&=&\frac{1}{(4\pi)^{d/2}}\frac{d}{2}\frac{\Gamma(n-d/2-1)}{\Gamma(n)}\frac{1}{\Delta^{n-d/2-1}},\nonumber
\end{eqnarray}
At first, let us consider the case of $\vec{k}_{\perp}=0$. A
straightforward calculation shows that
\begin{eqnarray}
\Pi(k_{0}=0,\vec{k}_{\perp}=0,\frac{k_{3}}{\sqrt{A}})
=-\frac{g^{2}}{8\pi^{2}\upsilon^{2}\sqrt{A}}\int^{\infty}_{-\infty}dq_{3}
\frac{2k_{3}^{2}q_{3}^{2}}{q_{3}^{2}+\frac{1}{4}k_{3}^{2}}. \nonumber
\end{eqnarray}
Since
the above integration is divergent, we introduce a UV cutoff $\Lambda_{3}$ for
$q_{3}$ variable. Then the resulting expression is given by
\begin{eqnarray}
\Pi(k_{0}=0,\vec{k}_{\perp}=0,\frac{k_{3}}{\sqrt{A}})
=-\frac{g^{2}}{8\pi^{2}\upsilon^{2}\sqrt{A}}[4\Lambda_{3}
k_{3}^{2}-\pi|k_{3}|^{3}], \nonumber
\end{eqnarray}
which shows that the leading $k_{3}$
dependence is still given by $k_{3}^{2}$. Similarly, we consider the case of
$k_{3}=0$, which is given by
\begin{eqnarray}
\Pi(k_{0}=0,\frac{\vec{k}_{\perp}}{\upsilon},k_{3}=0)
=-\frac{g^{2}}{8\pi^{2}\upsilon^{2}\sqrt{A}}\frac{3}{5}\pi k_{\perp}^{3/2}
=-\frac{3g^{2}}{40\pi\sqrt{\upsilon
A}}(\frac{k_{\perp}}{\upsilon})^{3/2}. \nonumber
\end{eqnarray}
Therefore $k_{\perp}^{3/2}$
dependence of the polarization function can be obtained consistent with the
result of RG analysis.

\newpage
\noindent{\bf Supplementary Note 5}
\\
{\bf Strong Coupling analysis.}

In this section, let us understand the
stability of the stable fixed point with the strong coupling analysis. The
strong coupling analysis starts with the following action,
\begin{eqnarray}
S_{\epsilon} &=& \int_{x,\tau} \psi^{\dagger}\left(\mathcal{H}_0 + i \varphi
\right) \psi +N_f \int_{q,\omega} (q_{\perp}^{2-\epsilon} +
q_3^2)|\varphi_{q,\omega}|^2, \nonumber
\end{eqnarray}
omitting unimportant
dimensionful numbers. The strong coupling analysis can be verified by the large
$N_f$ analysis introducing $N_f$ copies of different fermions. Here instead of
evaluating the exact ($N_f \rightarrow \infty$) result, we use the simplified
RPA results which capture the correct functional dependence. Since we are only
interested in how the infra-red divergence appears, the RPA result is enough to
estimate. One parameter ($\epsilon$) is introduced for comparison, and the
discussed anisotropic partial screening corresponds to $(\epsilon=1/2)$.

Given the  approximated action, we can evaluate $1/N_f$ correction.
The electron self energy with the momentum cutoff ($\Lambda$) in the quadratic
direction is
\begin{eqnarray}
\Sigma_{f}(k,\omega) &=& \frac{1}{N_f}\int d^3 \,q
\frac{\epsilon_a(k+q) \sigma^a}{E_{k+q}}
\frac{1}{q_{\perp}^{2-\epsilon}+q_3^2}\nonumber
\end{eqnarray}
where
$\epsilon_a \sigma^a = k_1 \sigma_1+k_2 \sigma_2+k_3^2 \sigma^3 $ and
$E(k)=\sqrt{(\epsilon^a(k))^2}$. It is easy to show that the correction from the
self-energy is not diverging in the infra-red cutoff,
\begin{eqnarray}
\frac{\delta \Sigma_{f}}{\delta \epsilon_a} \sim
\frac{1}{N_f}\int^{\Lambda}_{\mu}  \frac{d q_3}{q_3} q_3^{\epsilon/(2-\epsilon)}\sim
\Lambda^{\epsilon/(2-\epsilon)} \nonumber -\mu^{\epsilon/(2-\epsilon)},
\end{eqnarray}
for $\epsilon<1$. It is manifest that one can set $\mu \rightarrow 0$.
The absence of the infra-red divergence indicates the irrelevance
of the coulomb interaction.

One can further investigate the anisotropic fermion in two spatial dimensions.
Following the same method, one can show the linear direction is renormalized
significantly, so the strong coupling action becomes
\begin{eqnarray}
S_{2d} &=&
\int_{x,\tau} \psi^{\dagger}\mathcal{H}_2 \psi + i \varphi \psi^{\dagger} \psi
+\int_{q,\omega} (\sqrt{|q_{1}|} + |q_2|)|\varphi(q,\omega)|^2, \nonumber
\end{eqnarray}
where $\mathcal{H}_2(k) = k_1 \sigma^1 +k_2^2 \sigma^2 $. Then,
the correction from the fermion self energy is
\begin{eqnarray}
\frac{\delta
\Sigma_{f}}{\delta \epsilon_a} \sim \int d^2 q
\frac{1}{\sqrt{q_1^2+q_2^4}}\frac{1}{\sqrt{|q_1|}+|q_2|} \sim
\log(\frac{\Lambda}{\mu}) \nonumber
\end{eqnarray}
Thus, the anisotropic
screening induces the logarithmic correction similar to conventional graphene
physics.

%\begin{figure*}[t]
%\centering
%\includegraphics[width=16 cm]{Figure1.eps}
%\caption{
%{\bf Lattice structure and phase diagram of the 3D bulk system.}
%({\bf a}) Lattice structure of the pyrochlore lattice composed of
%corner sharing tetrahedrons.
%({\bf b}) Local spin structure of the AIAO antiferromagnet (AFM).
%({\bf c}) Schematic 3D bulk phase diagram as a function of local Coulomb interaction $U$.
%For $U>U_{c1}$, the system is an AIAO antiferromagnetic (AF) state.
%({\bf d}) Distribution of eight Weyl points for the Weyl-SM phase.
%All eight Weyl points are aligned along the [111] or its symmetry equivalent directions.
%Here red (blue) dot indicates a Weyl point with the chiral charge +1 (-1).
%} \label{fig:3Dlattice}
%\end{figure*}
%%%%%%%%%%%%%%%%%%%%%%%%%%%%%%%%%%%%%%%%%%%%%%%%%%%%%%%%%%%%%%%%%%%

\end{document}